\documentclass[a4paper,fleqn,usenatbib]{mnras}

\pdfoutput=1

\usepackage[T1]{fontenc}
\usepackage{ae,aecompl}

\usepackage{graphicx}	
\usepackage{amsmath}	
\usepackage{amssymb}	
\usepackage{comment}	
\usepackage{color}	

\title[Distance dependence and scaling relations]{Mutual distance dependence drives the observed jet power - radio luminosity scaling relations in radio galaxies.}

\author[L. E.H. Godfrey et al.]{
L. E.H. Godfrey,$^{1}$\thanks{E-mail: godfrey@astron.nl}
S. S. Shabala,$^{2}$
\\
$^{1}$ASTRON, the Netherlands Institute for Radio Astronomy, Postbus 2, 7990 AA, Dwingeloo, The Netherlands. \\
$^{2}$School of Physical Sciences, University of Tasmania, Private Bag 37, Hobart, Tas 7001, Australia
}

\date{Accepted 2015 November 17.  Received 2015 November 16; in original form 2015 September 17.}

\pubyear{2015}

\begin{document}
\label{firstpage}
\pagerange{\pageref{firstpage}--\pageref{lastpage}}
\maketitle

\begin{abstract}
The kinetic power of radio jets is a quantity of fundamental importance to studies of the AGN feedback process and radio galaxy physics. A widely used proxy for jet power is the extended radio luminosity. A number of empirical methods have been used to calibrate a scaling relationship between jet power (Q) and radio luminosity (L) of the form $\log(Q) = \beta_L ~ \log(L) + C$. The regression slope has typically been found to be $\beta_L \sim 0.7 - 0.8$. Here we show that the previously reported scaling relations are strongly affected by the confounding variable, distance. We find that in a sample of FRI X-ray cavity systems, after accounting for the mutual distance dependence, the jet power and radio luminosity are only weakly correlated, with slope $\beta_L \approx 0.3$: significantly flatter than previously reported. We also find that in previously used samples of high-power sources, no evidence for an intrinsic correlation is present when the effect of distance is accounted for. Using a simple model we show that $\beta_L$ is expected to be significantly lower in samples of FRI radio galaxies than it is for FRIIs, due to the differing dynamics for these two classes of radio source. For FRI X-ray cavity systems the model predicts $\beta_L ({\rm FRI}) \gtrsim 0.5$ in contrast to FRII radio galaxies, for which $\beta_L({\rm FRII}) \gtrsim 0.8$. We discuss the implications of our finding for studies of radio mode feedback, and radio galaxy physics.
\end{abstract}

\begin{keywords}
galaxies: active -- galaxies: jets -- radio continuum: galaxies
\end{keywords}


\section{Introduction}


The extended radio emission observed in radio galaxies arises from plumes or lobes of magnetised plasma that are deposited by powerful jets ejected from the galactic nucleus. Only a small fraction of the jet power is ultimately converted to synchrotron emission, with most of the jet energy being used to expand the lobes, drive shocks in the surrounding gas, and increase the amount of stored energy in the form of plasma internal energy and magnetic field \citep{bicknell97, hardcastle14}. The ratio of extended radio luminosity to jet power, $\frac{\nu L_\nu}{Q_{\rm jet}}$, which we shall call the extended radiative efficiency, varies greatly between sources. Measured at 1400 MHz, the extended radiative efficiency ranges from $\gtrsim 10^{-2}$ \citep[Cygnus A;][]{birzan08} to $\lesssim 10^{-5}$ \citep[MS0735.6+7421;][]{mcnamara05}. The extended radiative efficiency varies with the age of the source, and depends on the physical conditions within the lobes, as well as the evolutionary history of the radio galaxy \citep{bicknell97}, each of which depend on the external environment to some extent \citep[eg.][]{hardcastle13}. For this reason, the radio luminosity is not an accurate gauge of jet power in any individual object \citep{barthel96, hardcastle13}. 

However, the population of radio galaxies covers a range of more than 7 decades in radio luminosity, and on a population basis, the radio luminosity is expected to correlate strongly with the rate at which energy is deposited in the lobes. In principle, the scaling relation between radio luminosity and jet power can be a powerful tool in statistical studies of radio galaxies and their energetic impact on the surroundings, provided the scaling relation can be adequately calibrated. The $Q_{\rm jet} - L_\nu$ scaling relation is typically assumed to be a power-law of the form $Q_{\rm jet} = A L_{\nu}^{\beta_L}$ which is often written as $\log(Q_{\rm jet}) = \beta_L ~ \log(L_{\nu}) + Q_0$, where $Q_{\rm jet}$ is the jet power and $L_\nu$ is the monochromatic radio luminosity at frequency $\nu$. A lot of effort has been devoted to calibrating this scaling relation, from both a theoretical and empirical point of view, as we now discuss.

\citet{willott99} demonstrated a clear intrinsic correlation between the [OII] narrow line luminosity $L_{\rm NLR}$ and monochromatic radio luminosity $L_{\rm 151~MHz}$ with $L_{\rm NLR} \propto L_{\rm 151~MHz}^{0.79 \pm 0.04}$. Taking $L_{\rm NLR}$ as a proxy for accretion rate, and therefore jet power, \citet{willott99} argued that the observed $L_{\rm NLR} - L_{\rm 151~MHz}$ correlation provides an empirical estimate of $\beta_L$ for FRII radio galaxies. We further discuss the use of emission lines as a proxy for jet power, and in particular the shortcomings of such an approach, in Section \ref{sec:emission_line_proxies}.

\citet{willott99} also presented a model-dependent predictor of jet power based on synchrotron minimum energy calculations in combination with the self-similar model of radio galaxy evolution \citep{falle91, kaiser97}. They obtain an expression for the jet power $Q_W$ (``W" for Willott) in terms of the 151 MHz radio luminosity
\begin{equation} \label{eqn:willott}
Q_W \approx f^{3/2} ~ 3 \times 10^{38} \left(  \frac{L_{151}}{10^{28}~{\rm W~Hz^{-1}~sr^{-1}}} \right)^{6/7} ~ {\rm W}
\end{equation}
where $Q_W$ is time averaged kinetic power of a source with radio luminosity $L_{151} = F_{\rm 151} D_L^2$, where $D_L$ is the luminosity distance and $f$ is a parameter accounting for systematic error in the model assumptions. These model assumptions include, among other things, the fraction of energy in non-radiating particles, the low frequency cutoff in the synchrotron spectrum, and departures from minimum energy. It is argued by \citet{willott99} that $1 \leq f \leq 20$, implying a systematic uncertainty of 2 orders of magnitude in jet power for a given radio luminosity, owing to the $f^{3/2}$ dependence. Despite this uncertainty, the expression for $Q_W$ above is widely used to estimate the mechanical output from AGN based on a single low frequency luminosity measurement, assuming that the value of $f$ is constant (typically of order 10 - 20) across the entire population of radio galaxies \citep[e.\,g.][]{hardcastle07, martinez-sansigre11, fernandes11, cattaneo09}.  

A number of empirical jet power measurement techniques have been employed to calibrate the $Q_{\rm jet} - L_\nu$ scaling relation, and test the validity of Equation \ref{eqn:willott}. Arguably the most direct measure of AGN mechanical power is the so-called X-ray cavities method (see Section \ref{sec:cavities} and references therein). This method relies on the detection of X-ray surface brightness depressions (X-ray cavities) associated with the extended radio lobes. The jet power is calculated based on the energy required to inflate the cavities. X-ray cavity power measurements are inherently limited to systems in which X-ray cavities can be clearly detected: that is, relatively nearby low power objects in dense environments \citep{mcnamara12}, typically of Fanaroff-Riley type I morphology \citep[FRI;][]{fanaroff74}. \citet{birzan08} presented an analysis of jet power measurements for a sample of 24 X-ray cavity systems predominantly in galaxy clusters, and found $\beta_L \approx 0.5 - 0.7$, albeit with a very large intrinsic scatter in the relation. \citet{cavagnolo10} extended the sample of \citet{birzan08} to lower jet power by adding several X-ray cavity systems associated with giant Elliptical galaxies to the sample, and found $\beta_L \approx 0.7$. \citet{cavagnolo10} argued that their empirical relation is consistent with the model of \citet{willott99} (Equation \ref{eqn:willott}), provided that the energy density of non-radiating particles in the lobes is 100 times that of the relativistic electron population. \citet{osullivan11} confirmed the results of \citet{cavagnolo10}, by similarly extending the sample of \citet{birzan08}. 

\citet{daly12} estimated the jet power for a sample of 31 high power FRII radio galaxies, using the expression $Q = 4 p V / \tau$, where $p$ is the lobe pressure calculated using minimum energy arguments, $V$ is the lobe volume assuming cylindrical symmetry, and $\tau$ is the spectral age of the source. Using this method, \citet{daly12} found the scaling relation between jet power and radio luminosity in their sample to be consistent with the model prediction of \citet{willott99} (Equation \ref{eqn:willott}), and also consistent with an extrapolation of the scaling relation for FRI radio galaxies. 

These empirical calibrations of the $Q_{\rm jet} - L_\nu$ scaling relation appear to be in broad agreement with each other, and also appear to support the model predictions of \citet{willott99}. However, \citet[][GS13]{godfrey13} pointed out that due to the large difference in energy budget and dynamics of FRI and FRII radio galaxy lobes, the apparent agreement in the FRI and FRII scaling relations is entirely unexpected. It was argued by GS13 that the $Q_{\rm jet} - L_{\nu}$ scaling relations should differ greatly between the two classes of object, in both slope and normalisation. In an effort to test this hypothesis, GS13 devised a new method for measuring jet power in FRII radio galaxies based on the observed hotspot parameters. Using this new method with a sample of 29 3C FRII radio galaxies, it was found that the $Q_{\rm jet} - L_\nu$ scaling relation agreed with the model predictions of \citet{willott99} as well as the empirical results of \citet{daly12} for FRII radio galaxies, and was strikingly similar to that obtained for FRI radio galaxies by \citet{cavagnolo10}, despite expectations to the contrary. 

The agreement between the scaling relations in GS13 appeared to confirm the previously held position that $Q_W$ (Equation \ref{eqn:willott}) could be applied to the entire radio galaxy population, regardless of source morphology, environment, or jet power. However, this conclusion was erroneous: here we show that the apparent agreement between the various empirically derived scaling relations is due to the similar distance dependence of jet power measurement techniques used for FRI and FRII radio galaxies. In each case described above, except for the study of \citet{willott99}, the effect of distance has been neglected. The purpose of this paper is to present a re-analysis of the previously reported scaling relations for FRI and FRII radio galaxies, accounting for the distance dependence in jet power measurements.  

As already mentioned, there is a large intrinsic scatter in the relationship between jet power and radio luminosity from source to source. Therefore, to enable precise calibration of the average scaling relation, it is necessary to use samples that cover a broad range in luminosity and jet power. In the case of \citet{cavagnolo10}, for example, the sample covers 6 dex in radio luminosity and 5 dex in jet power. However, to cover a such broad range in physical parameters, the sample necessarily spans a very wide range in distance. The samples are therefore subject to Malmquist bias. When spanning a large range in distance, this can potentially result in a spurious relationship between jet power and radio luminosity, which is driven by the common distance dependence on both axes in the sample \citep[eg.][]{feigelson83}. This spurious relation can dominate over any intrinsic relationship between the variables, or can produce a strong apparent correlation when no intrinsic relation exists. 

In Section \ref{sec:cavities} we consider the scaling relation for FRI radio galaxies based on X-ray cavity jet power measurements. 
In section \ref{sec:FRIIs} we consider the scaling relation for FRII radio galaxies, derived using various different measurement techniques. 
In Section \ref{sec:model_predictions}, we derive a model for the $Q_{\rm jet} - L_\nu$ scaling relation in different types of radio galaxy, and make a comparison between model predictions and the observed regression slope. 
In Section \ref{sec:conclusions} we summarise our findings and further consider the implications for mechanical feedback (AGN feedback by the radio jets), and radio galaxy physics. 

\section{Jet power from X-ray Cavities}  \label{sec:cavities}

X-ray images of hot gaseous halos surrounding massive galaxies that host radio AGN sometimes show depressions in the X-ray surface brightness associated with regions of extended radio emission \citep{boehringer93}. These X-ray surface brightness depressions are interpreted as bubbles or ``cavities" in the hot gas that are created by the expansion of the radio lobes in the surrounding gas \citep{mcnamara00, blanton01, johnstone02, dunn04}. The amount of work required to create the observed cavities provides a measure of the total mechanical energy produced by the AGN. This estimate of total mechanical energy, when combined with an estimate of the age of the radio galaxy outburst, provides a measurement of the time averaged jet power \citep{mcnamara00, churazov02, birzan04}: 
\begin{equation}
\label{eqn:cavity_power_1}
Q_{\rm jet} = \frac{W}{\tau} \approx \frac{\zeta p V}{\tau}.
\end{equation}
The pre-factor $\zeta$ depends on the equation of state of the plasma within the cavity, as well as the expansion history of the bubble, but is typically taken to be $\zeta = 4$ \citep{mcnamara12}. The pressure, $p$, is determined from X-ray imaging spectroscopy at the mid-point of the cavity, the volume $V$ is determined from the angular extent of the cavity (assuming eg. ellipsoidal geometry), and the age of the system $\tau$ is estimated by a variety of means, but is typically taken to be the buoyancy timescale for the bubble to rise to its current height in the hot atmosphere \citep[eg.][]{rafferty06, birzan08, cavagnolo10, osullivan11}. The jet power estimated in this way is often called the cavity power.

The X-ray cavity method has been applied to a wide range of systems covering a wide range of cavity power, from sources in clusters and groups to massive early type galaxies \citep[see the review of][]{mcnamara12}. By comparing the cavity power and radio luminosity in a large sample of X-ray cavity systems, it is possible to determine a scaling relationship between jet power and radio luminosity \citep{birzan04, birzan08, cavagnolo10, osullivan11}. However, the sample necessarily consists entirely of systems with detected X-ray cavities, which introduces a strong selection bias that is manifested as a tight relationship between jet power and distance in the sample. In Section \ref{sec:expected_cavities_distance_dependence} we derive the distance dependence of cavity power measurements. In Section \ref{sec:cavities_sample} we discuss the X-ray cavities sample and the uncertainty calculations. In Section \ref{sec:cavities_regression} we present our data analysis and results. Finally, in Section \ref{sec:model_predictions} we derive a model to determine the expected scaling relation for FRI X-ray cavity systems.

\subsection{Distance dependence of X-ray cavity jet power measurements} \label{sec:expected_cavities_distance_dependence}

In order to demonstrate the distance dependence of the X-ray cavity jet power measurements, we consider the simple case of a cavity in an isothermal gas at temperature $T$ with a $\beta$-model density distribution (King atmosphere) of the form
\begin{equation}  \label{eqn:density_profile}
\rho = \rho_0 \left( 1 + \frac{r^2}{r_c^2}  \right)^{-\frac{3}{2} b}
\end{equation}
where $r_c$ is the core radius and $\rho_0$ is the central density that provides the normalisation for the density distribution, and $b$ is the exponent of the power-law density profile. 
For a given temperature, the X-ray surface brightness $\Sigma$ as a function of angular distance from the centre can be written as \citep[][]{birkinshaw93, worrall06}:
\begin{equation}
\Sigma_\nu (\theta) \propto n_{e}(\theta) n_{p}(\theta) \frac{D_A}{(1+z)^3}   \left(1 + \frac{\theta^2}{\theta^2_c}  \right)^{\frac{1}{2}}
\end{equation}
where $n_{e}(\theta)$ and $n_{p}(\theta)$ are the electron and proton density at a clustercentric radius $r = D_A \theta$, respectively, and $D_A$ is the angular diameter distance. The product $n_e n_p \propto p^2$ where $p$ is the gas pressure, and so at a fixed surface brightness, temperature, and clustercentric radius we can write 
\begin{equation}
p \propto (1+z)^{3/2} D_A^{-1/2} \label{eqn:p_D_A}
\end{equation}
For a given angular size of a cavity, the volume is  
\begin{equation}
V \propto D_A^{3} \label{eqn:V_D_A}
\end{equation}
The age $\tau$ is typically estimated as the time required for the bubble to rise buoyantly at the terminal velocity $v_t$ to it's current radius, $R$, and is calculated as
\begin{equation}
\tau_{\rm buoy} = R/v_t = R \left( \frac{S~C}{2gV} \right)^{1/2}  
\end{equation} 
where S is the cavity cross-section, $C \approx 0.75$ is the drag coefficient, V is the volume, and $g$ is the gravitational acceleration which is calculated using the stellar velocity dispersion of the host galaxy under the approximation $g \approx 2 \sigma^2/R$ \citep[see][]{churazov01, birzan04}. Hence,
\begin{equation}
\tau_{\rm buoy} = R \left( \frac{S~C~R}{4 \sigma^2 V} \right)^{1/2} \propto D_A. \label{eqn:t_buoy_D_A}
\end{equation} 
From equations \ref{eqn:cavity_power_1}, \ref{eqn:p_D_A}, \ref{eqn:V_D_A} and \ref{eqn:t_buoy_D_A} we find
\begin{equation}
Q_{\rm jet} \propto D_A^{3/2} (1+z)^{3/2} \propto D_L^{3/2} (1+z)^{-3/2}.
\end{equation} 
Samples of X-ray cavity systems typically include only low redshift radio galaxies, so the factor $(1+z)^{-3/2}$ has a small effect on the distance dependence. Therefore, given a sample of objects with a narrow range of observational properties (i.e. X-ray surface brightness distribution, cavity angular size, cavity angular offset from the centre of the hot atmosphere, and radio flux density) relative to the range in distance squared, the distance dependence of $Q_{\rm jet} \sim D_L^{1.5}$ and $L_\nu \sim D_L^2$ will induce a spurious relationship $Q_{\rm jet} \sim L_\nu^{0.75}$. The narrow range in the observed parameters result from various selection effects in combination with the steep radio and kinetic luminosity functions. Some of the selection effects are not obvious, for example, the scale of the cavities cannot be very large relative to the core radius in the density profile ($r_c$ in Equation \ref{eqn:density_profile}), due to the lack of signal to noise for a detection of the X-ray cavity \citep{ensslin02}. In contrast, the range in distance covered by the sample is very large, spanning more than two orders of magnitude (see Figure \ref{fig:cavities_327_data_plots}).

The predicted spurious relation is similar to the observed scaling relation $Q_{\rm jet} \sim L_\nu^{0.7}$ \citep[eg.][]{cavagnolo10, osullivan11}. In the following sections we investigate the importance of the distance dependence, and its effect on the observed scaling relation.

\subsection{The X-ray cavities sample} \label{sec:cavities_sample}

For the following analysis, we combine the samples of \citet{birzan08}, \citet{cavagnolo10} and \citet{osullivan11}. Since we do not expect FRI and FRII radio galaxies to follow the same relationship \citep[see][]{godfrey13}, we exclude Cygnus A. The samples of \citet{cavagnolo10} and \citet{osullivan11} share several sources in common\footnote{The samples of \citet{cavagnolo10} and \citet{osullivan11} also have in common the sample of \citet{birzan08}, but here we specifically mean sources in addition to the \citet{birzan08} sample that are in common between the two studies.}. For these sources, we use the distances, luminosities and cavity powers taken from \citet{osullivan11} due to the improved radio data available in that study. We note, however, that the cavity power estimates of \citet{cavagnolo10} typically agree with those of \citet{osullivan11} to within a factor of 1 -- 3. 

For sources at $D_L > 70$~Mpc we use redshift derived distance estimates and assume $\Delta D_L = 7$~Mpc corresponding to peculiar velocities of $\sigma_{\rm v} \approx 500$~km/s. For nearby elliptical galaxies at $D \lesssim 40$~Mpc, redshift independent distance measurements have a typical accuracy of order 10$\%$ - 20$\%$, depending on the method used \citep{cappellari11}. Therefore, for sources with $D_L < 70$~Mpc we assume $\Delta D_L = 0.1 D_L$, coresponding to the estimated uncertainty in redshift independent distance measurements. The only exception to these rules are M84 and M87, for which distances have been measured using the surface brightness fluctuation method with the Hubble Space Telescope Advanced Camera for Surveys (ACS), and are deemed to be accurate to $\sim 3 \%$ \citep{blakeslee09}. 

Uncertainties in distance are propagated when calculating the uncertainties in luminosity,
\begin{equation}
\Delta L_\nu = L_\nu ~ \sqrt{\left(  \frac{2 \Delta D_L}{D_L} \right)^2 + \left( \frac{\Delta S_\nu}{S_\nu} \right)^2}
\end{equation}
where $S_\nu$ is the radio continuum flux density. As a result, the uncertainties in luminosity are greater than those quoted in the original studies, and are correlated with the distance uncertainties. We do not propagate the distance uncertainties with jet power, since the jet power uncertainties are strongly dominated by other sources of error such as volume estimate \citep{osullivan11}, and the added contribution due to uncertainty in distance can be safely neglected.

\subsection{Partial correlation analysis and Bayesian multivariable linear regression: accounting for the distance dependence}  \label{sec:cavities_regression}

\begin{figure*}
   \includegraphics[width=2.0\columnwidth]{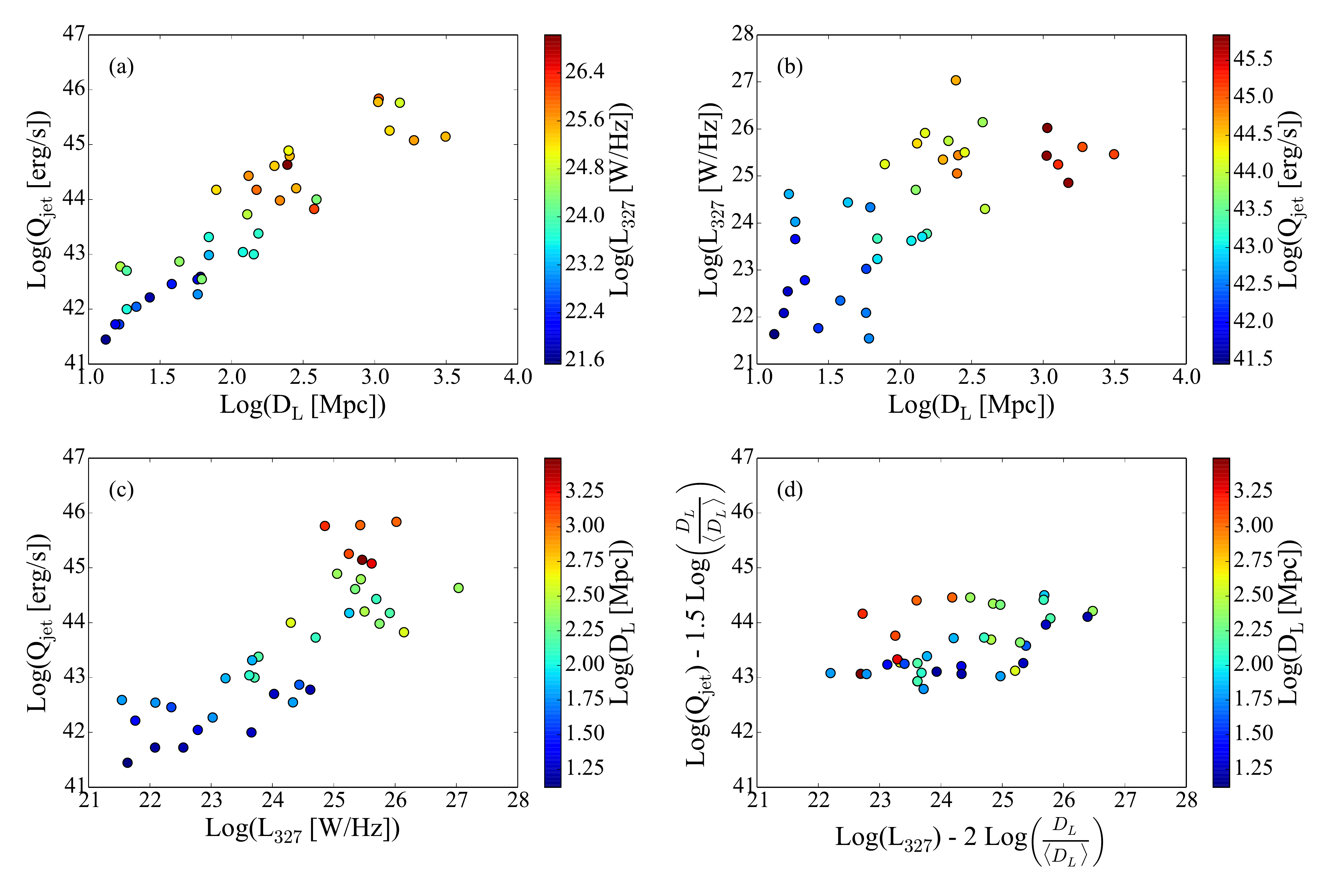}
    \caption{The interdependence of jet power ($Q_{\rm jet}$), radio luminosity at 327 MHz ($L_{\rm 327}$) and luminosity distance ($D_L$) for the combined X-ray cavities sample of \citet{birzan08}, \citet{cavagnolo10} and \citet{osullivan11} (see Section \ref{sec:cavities_sample}). In each 2D panel, the third axis is represented by the colour scale. In this sample, the tightest correlation is between jet power and distance. The lower right panel is the distance normalised plot, equivalent to a flux-flux plot that is often used in the analysis of correlations between luminosities in different wavebands. The distance normalised jet power is obtained by dividing the jet power by the expected distance dependence derived in Section \ref{sec:expected_cavities_distance_dependence} ($Q_{\rm jet} \sim D_L^{1.5}$) and of course luminosity scales as $L_\nu \propto D_L^2$ . The distance dependence is scaled by the median distance $\langle D_L \rangle$ and plotted on the same axes scale as the un-normalised $Q_{\rm jet} - L_\nu$ plot, in order to demonstrate the degree of ``stretching" in each axis due to the broad range of distance. Comparison of panels (c) and (d) shows that the range of jet power spans 5 dex, while the distance normalised jet powers span only 2 dex. The distance normalised plots are not used to determine the intrinsic scaling relation, and are included here for illustrative purposes only. A full multivariable linear regression is used to disentangle the distance and luminosity dependence of jet power. See the electronic edition of the Journal for a colour version of this figure.}
    \label{fig:cavities_327_data_plots}
\end{figure*}

\begin{figure*}
   	\includegraphics[width=2.0\columnwidth]{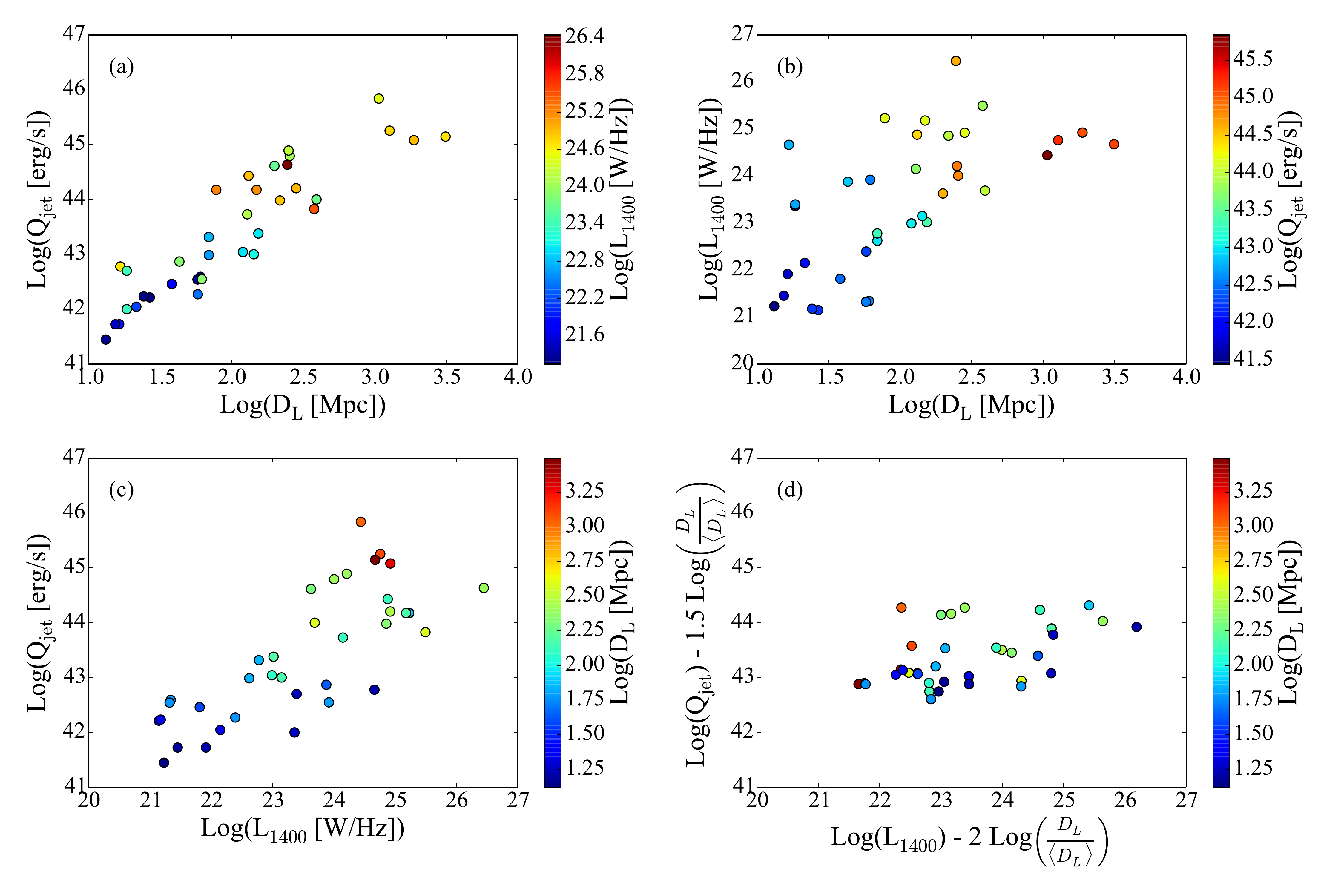}
    \caption{The interdependence of jet power ($Q_{\rm jet}$), radio luminosity at 1400 MHz (L$_{1400}$) and distance ($D_L$) for the combined X-ray cavities sample of \citet{birzan08}, \citet{cavagnolo10} and \citet{osullivan11} (see Section \ref{sec:cavities_sample}). Panels are the same as Figure \ref{fig:cavities_327_data_plots}. See the electronic edition of the Journal for a colour version of this figure.}
    \label{fig:cavities_1400_data_plots}
 \end{figure*}

\begin{figure*}
	\includegraphics[width=1.6\columnwidth]{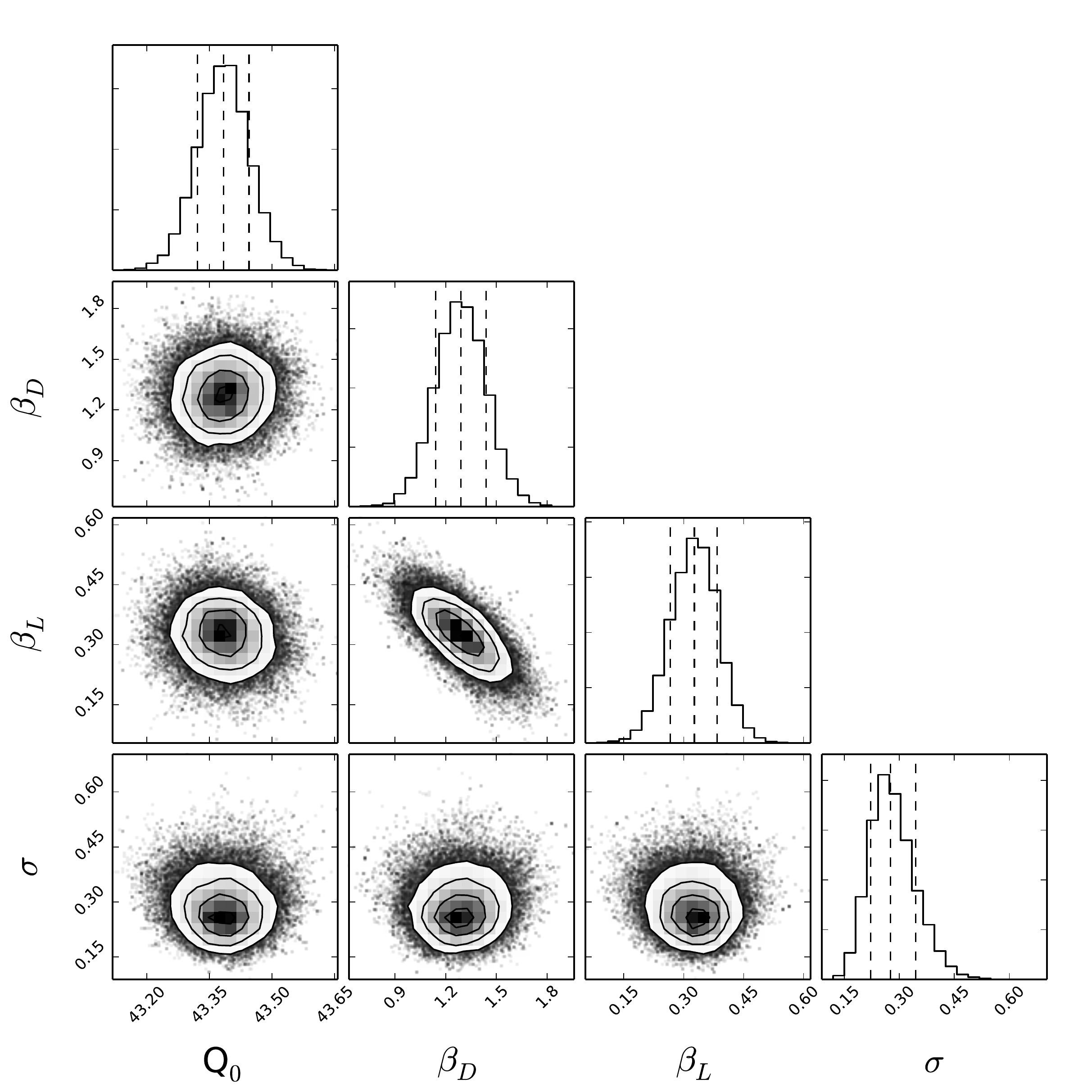}
    \caption{Posterior probability distributions for the multivariable linear regression model parameters for the X-ray cavities sample at 327 MHz. The model parameters $Q_0, \beta_D, \beta_L$ and $\sigma$ are defined by Equation \ref{eqn:regression_model}.}
  \label{fig:cavities_corner_plot}
 \end{figure*}

\begin{figure}
	\includegraphics[width=\columnwidth]{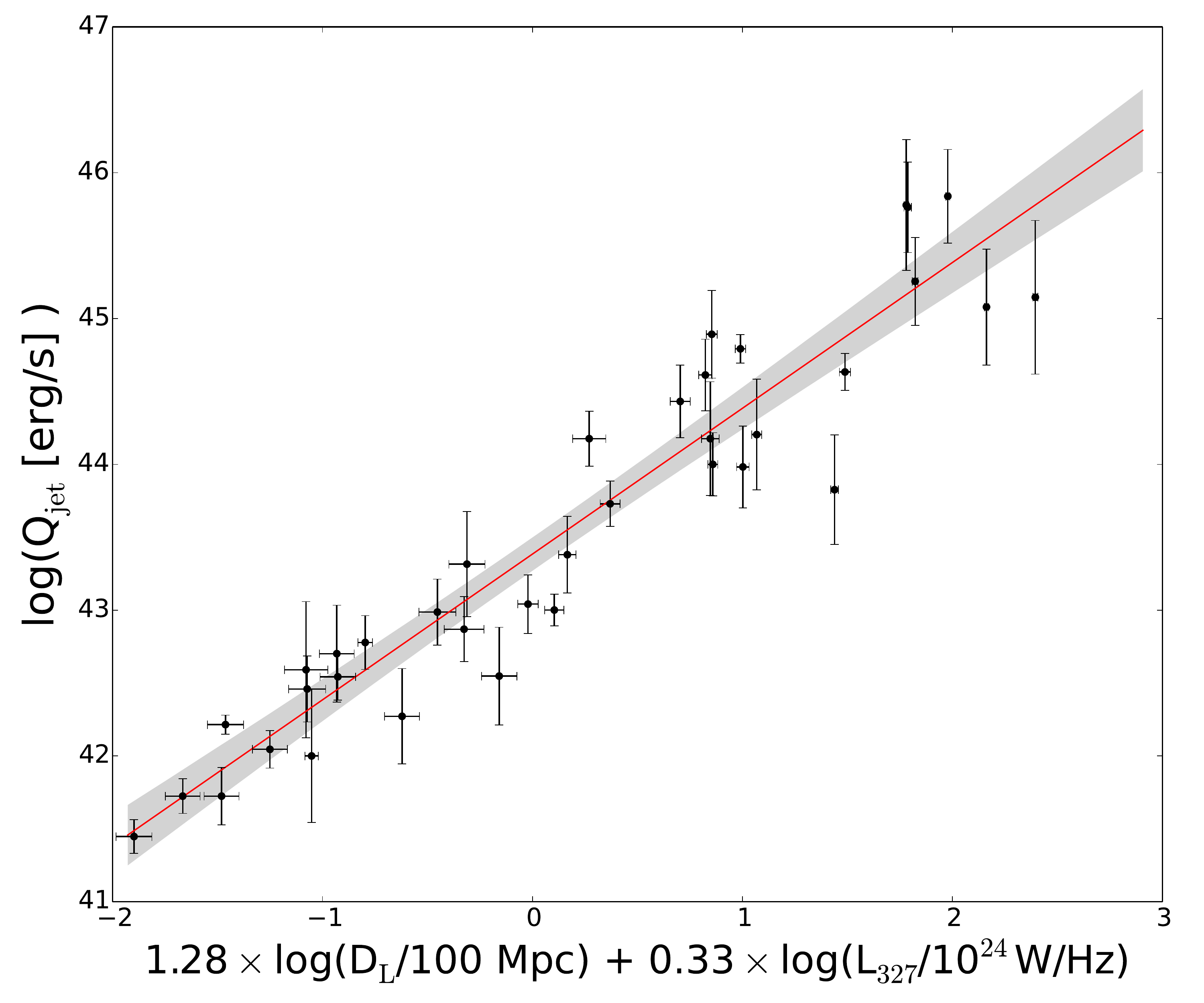}
    \caption{The best-fit multivariable linear regression relation for the X-ray cavities sample at 327 MHz. The shaded area encompasses the $90\%$ confidence region. The tight correlation shown here is driven largely by the tight correlation between jet power and distance in this sample. The intrinsic scatter about the best-fit is $\sigma = 0.28^{+0.06}_{-0.05}$ dex, which is less than half the intrinsic scatter in the relation between jet power and luminosity, which has intrinsic scatter $\sigma \approx 0.6$ dex \citep{osullivan11}.}
   \label{fig:cavities_best_fits}
\end{figure}

\begin{figure*}
   	\includegraphics[width=2.0\columnwidth]{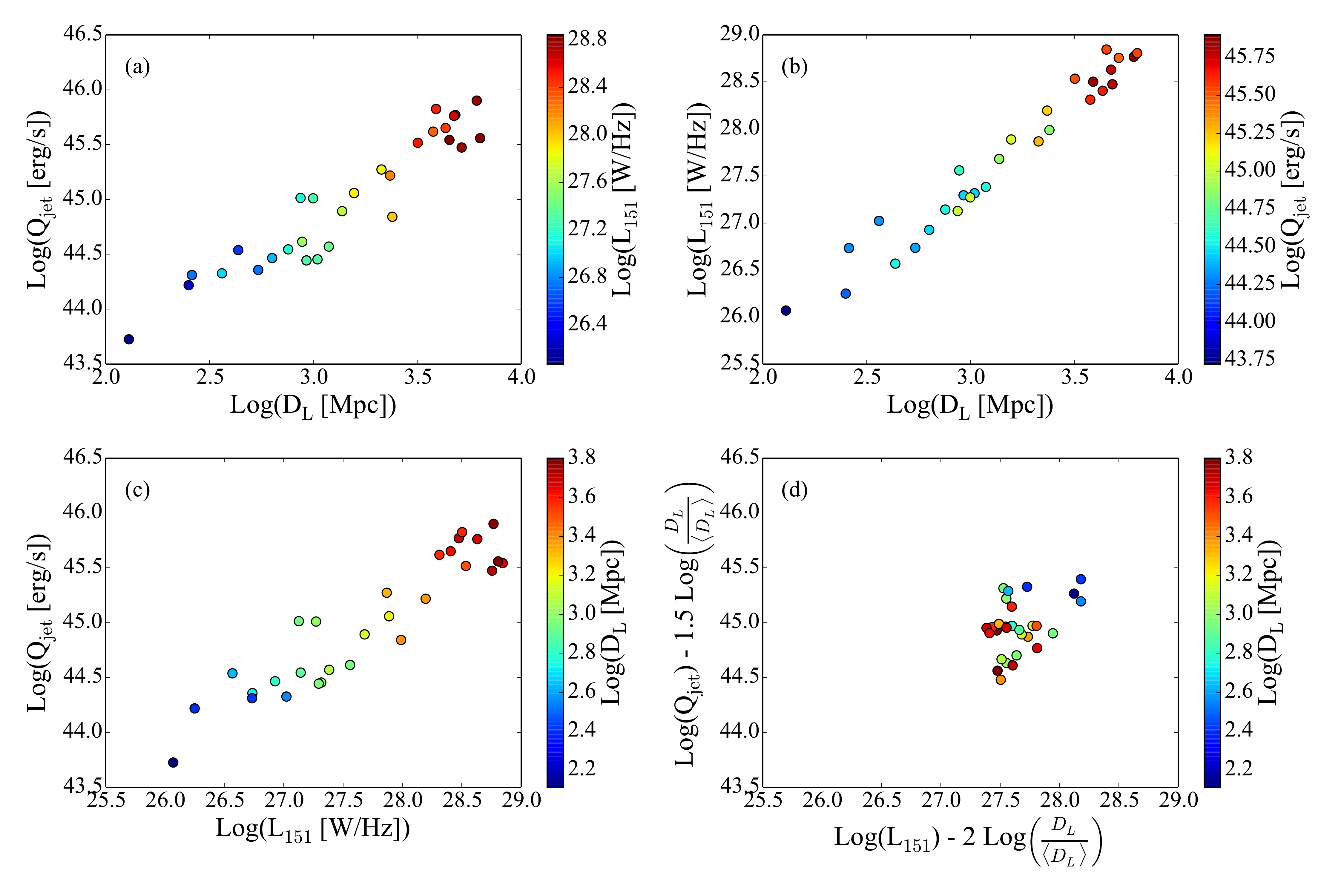}
    \caption{The interdependence of jet power ($Q_{\rm jet}$), radio luminosity at 151 MHz (L$_{151}$) and distance ($D_L$) for the FRII hotspot jet power measurements of \citet{godfrey13}. Panels are the same as Figure \ref{fig:cavities_327_data_plots}. The distance normalised jet power is obtained by dividing the jet power by the expected distance dependence of hotspot jet power measurements derived in Section \ref{sec:hotspots} ($Q_{\rm jet} \sim D_L^{1.5}$). The distance normalised plot shows the relatively narrow range spanned in both flux density and distance normalised jet power. This narrow range is the reason that no intrinsic correlation is obtained after accounting for the common distance dependence. See the electronic edition of the Journal for a colour version of this figure.}
    \label{fig:hotspots_data_plots}
 \end{figure*}

\begin{figure*}
   	\includegraphics[width=2.0\columnwidth]{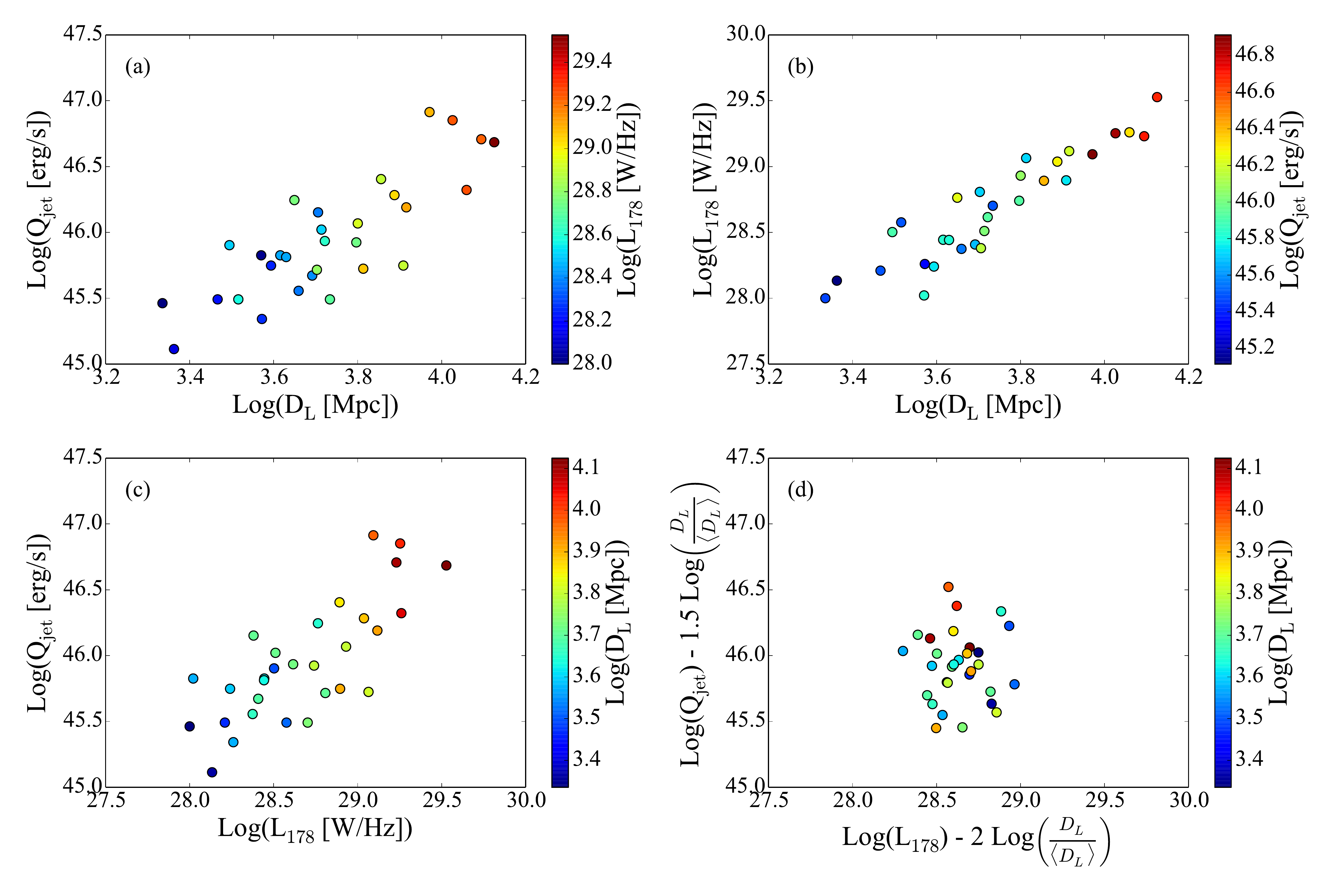}
    \caption{The interdependence of jet power ($Q_{\rm jet}$), radio luminosity at 178 MHz (L$_{178}$) and distance ($D_L$) for the FRII sample of \citet{daly12} (see Section \ref{sec:daly_sample}). Panels are the same as Figure \ref{fig:cavities_327_data_plots}. The distance normalised jet power is obtained by dividing the jet power by the expected distance dependence of jet power measurements based on minimum energy and spectral age, derived in Section \ref{sec:daly_sample} ($Q_{\rm jet} \sim D_L^{1.5}$). The distance normalised plot shows the relatively narrow range spanned in both flux density and distance normalised jet power. This narrow range is the reason no intrinsic correlation is obtained after accounting for the common distance dependence. See the electronic edition of the Journal for a colour version of this figure.}
    \label{fig:daly_data_plots}
 \end{figure*}

\begin{table}
 \caption{Results of partial correlation analysis for log($Q_{\rm jet}$), log($L_\nu$) and log($D_{\rm L}$).}
 \label{table:partial_correlation_analysis_results}
  \begin{tabular}{ccccc}
  \hline
Sample & $\tau_{\rm QD}$ & $\tau_{\rm LD}$ & $\tau_{\rm QL}$ & $\tau_{\rm QL|D}^{a}$ \\
  \hline
X-ray Cavities 327 MHz & 0.77 & 0.54 & 0.59 & $0.34 \pm 0.1$  \\
X-ray Cavities 1.4 GHz  & 0.77 & 0.50 & 0.56 & $0.31 \pm 0.1$ \\
GS13 FRII sample                        & 0.78 & 0.88 & 0.72 & $0.14 \pm 0.1$ \\
Daly et al. FRII sample                  & 0.57 & 0.77 & 0.56 & $0.24 \pm 0.13$ \\
  \hline
  \multicolumn{5}{l}{$^a$ $\tau_{QL|D}$ is the partial Kendall's $\tau$ correlation coefficient between}\\
\multicolumn{5}{l}{log($Q_{\rm jet})$ and log$(L_\nu)$ after the influence of the third variable,} \\
\multicolumn{5}{l}{log($D_L$), is accounted for. The partial Kendall's $\tau$ is } \\
\multicolumn{5}{l}{calculated using the Fortran code of \citet{akritas96}.} \\
 \end{tabular}
\end{table}

We begin by considering the mutual correlations between the key quantities. Figure \ref{fig:cavities_327_data_plots} shows that the jet power (Q), distance (D) and Luminosity (L) are correlated with each other. To quantify the interdependence of the three parameters we perform partial correlation analysis (see Table \ref{table:partial_correlation_analysis_results}) using Kendall's $\tau$ rank correlation coefficient \citep{akritas96}. Table \ref{table:partial_correlation_analysis_results} and Figure \ref{fig:cavities_327_data_plots} show that, for this sample of X-ray cavity systems, the distance to the object is a better predictor of jet power than the radio luminosity. \\

In panel d of Figure \ref{fig:cavities_327_data_plots} we present a distance normalised plot of jet power and luminosity, which is equivalent to a flux-flux plot that is often used in the analysis of correlations between luminosities in different wavebands (note, however, that the distance normalised plot is included for illustrative purposes only. A full multivariable linear regression is used to disentangle the distance and luminosity dependence of jet power). The distance normalised jet power is obtained by dividing the jet power by the expected distance dependence derived in Section \ref{sec:expected_cavities_distance_dependence} ($Q_{\rm jet} \sim D_L^{1.5}$). The distance dependence is scaled by the median distance $\langle D_L \rangle$ and plotted on the same axes scale as the un-normalised $Q_{\rm jet} - L_\nu$ plot, in order to demonstrate the degree of ``stretching" in each axis due to the broad range of distance. Comparison of panels (c) and (d) shows that the range of jet power spans 5 dex, while the distance normalised jet powers span only 2 dex. We note that the range in distance normalised jet power is comparable to the scatter in the relation.

We perform multivariable linear regression adopting a Bayesian approach implemented in Python using the affine invariant MCMC ensemble sampler ``emcee" \citep{foreman-mackey13}. We use an error-in-variables regression model that accounts for the covariant uncertainties in the independent variables, as well as uncertainty in jet power and intrinsic scatter in the model. To simplify the analysis, we assume Gaussian uncertainties in the jet power measurements, where the standard deviation in log(Q) is calculated as the average of the positive and negative quoted uncertainties: $\sigma_{\rm \log Q} = \frac{1}{\ln(10)} \times \left( \frac{\Delta Q_+ + \Delta Q_-}{2Q} \right)$.  We fit a model of the form
\begin{equation} \label{eqn:regression_model}
\log(Q_{\rm jet}) = Q_0 + \beta_L \log \left(\frac{L}{\rm 10^{24} W/Hz}\right) + \beta_{D} \log \left(\frac{D_L}{\rm 100 Mpc} \right) + \epsilon.
\end{equation}
where $\epsilon$ is a Gaussian error term with standard deviation $\sigma$ accounting for the intrinsic scatter in the relation. We assume priors on the regression coefficients $p(Q_0, \beta_L, \beta_D, \sigma^2) \propto (1+ \beta_L^2 + \beta_D^2)^{1/2}$, so that the model is rotationally invariant \citep[see][]{hogg10, robotham15}.

The results of the regression analysis are presented in Table \ref{table:regression_results}, Figure \ref{fig:cavities_corner_plot} and Figure \ref{fig:cavities_best_fits}. The relation between jet power and radio luminosity is significantly weaker than previously reported. The results of our multivariate regression indicate that jet power only gradually increases with increasing radio luminosity: the regression slope is $\beta_L = 0.33 \pm 0.09$ at 327~MHz and $\beta_L = 0.27 \pm 0.09$ at 1400~MHz. The odds ratio that $\beta_L < 0.5$ is greater than 400:1. This contrasts with the previously reported scaling relation for X-ray cavity systems using a similar sample, in which $\beta_L = 0.64 \pm 0.09$ \citep{cavagnolo10} and $\beta_L = 0.7 \pm 0.1$ \citep{osullivan11}. The flat regression slope we find in this work implies that a greater amount of mechanical energy is available from lower luminosity radio galaxies than previously thought. The implications are further discussed in Section \ref{sec:conclusions}.

\begin{table*}
 \caption{Results of multivariate regression: $\log(Q_{\rm jet}) = Q_0 + \beta_D \log(D_L/D_0)  + \beta_L \log(L_\nu/L_0) \pm \sigma$ (see equation \ref{eqn:regression_model}).}
\label{table:regression_results}
 \begin{tabular}{ccccccc}
  \hline
Sample & $Q_0$ & $\beta_D$ & $D_0$ [Mpc] &  $\beta_L$ & $L_0$ [W/Hz] & $\sigma$ \\
 & Normalisation & distance slope & pivot distance & luminosity slope & pivot luminosity & intrinsic scatter \\
  \hline
X-ray Cavities 327 MHz & $43.39 \pm 0.06$ & $1.3 \pm 0.2$ & $10^{2}$      &  $0.33 \pm 0.09$ & $10^{24}$ & $0.28_{-0.05}^{+0.06}$ \\
X-ray Cavities 1.4 GHz  & $43.56 \pm 0.08$ & $1.4 \pm 0.25$ & $10^{2}$      &  $0.27 \pm 0.09$ & $10^{24}$ & $0.33_{-0.06}^{+0.07}$ \\
Daly et al. FRII sample  & $45.91 \pm 0.06$ & $1.8 \pm 1.0  $   & $10^{3.7}$ & $0.0 \pm 0.5$      & $10^{28.7}$   & $0.25 \pm 0.05$ \\
GS13 FRII sample         & $45.02 \pm 0.06$ & $1.1 \pm 0.6  $   & $10^{3.2}$ &  $0.1 \pm 0.4$     & $10^{27.6}$   & $0.19 \pm 0.04$ \\
  \hline
    \multicolumn{7}{l}{Due to the strong correlation between $\beta_D$ and $\beta_L$, we determine uncertainties for these parameters based on the 2D 68.3$\%$ credible} \\
    \multicolumn{7}{l}{region, marginalising over the other model parameters. For $Q_0$ and $\sigma$, the uncertainties are based on 1D 68.3$\%$ credible intervals.} \\    
    \multicolumn{7}{l}{We again stress that the distance dependence in the regression relations account for selection effects within each of the samples,}\\
    \multicolumn{7}{l}{and therefore, the above distance dependence does not apply to the radio galaxy population as a whole.}\\
 \end{tabular}
\end{table*}

\section{High power (FRII) radio galaxies} \label{sec:FRIIs}

Due to the greater distance to powerful FRII radio galaxies, detection of X-ray cavities associated with powerful FRIIs is difficult. There are only two genuine FRII radio galaxies with robust X-ray cavity jet power measurements: Cygnus A \citep{wilson06} and 3C~444 \citep{croston11}. For this reason, alternative jet power measurement techniques have been sought. In the following sections we discuss two of these alternative methods, and perform the same correlation analysis as used in the previous section. 

\newpage

\subsection{Hotspot Jet Power Measurement}  \label{sec:hotspots}

\citet{godfrey13} devised a method for jet power measurement based on the observed parameters of the hotspots, and applied this method to a sample of FRII radio galaxies to determine the FRII $Q_{\rm jet} - L_\nu$ scaling relation. With this method, the jet power is given by 
\begin{eqnarray}
Q &\propto& A B_{\rm eq}^2 
\end{eqnarray}
where A is the hotspot cross-sectional area, and $B_{\rm eq}$ is the equipartition magnetic field strength in the hotspot (note that $B_{\rm eq}$ is related to the minimum energy magnetic field strength $B_{\rm me}$ via a function of the spectral index that is of order unity). Collecting terms involving distance, we find
\begin{eqnarray}
Q &\propto& D_L^{2 - \frac{2}{3 + \alpha}} \sim D_L^{1.5}
\end{eqnarray}
This is similar to the distance dependence of the X-ray cavity jet power measurement technique, and as with the X-ray cavity powers, it results in an expected spurious relation of the form $Q_{\rm jet} \sim L_\nu^{0.75}$. We note that, using the hotspot method, GS13 find $Q_{\rm jet} \sim L_\nu^{0.7}$.

Partial correlation analysis on the sample of GS13 indicates that the observed correlation between $Q_{\rm jet}$ and $L_\nu$ is indeed dominated by the mutual distance dependence (Table \ref{table:partial_correlation_analysis_results}), and this is confirmed by the results of our multivariable linear regression (Table \ref{table:regression_results}), in which we find a slope $\beta_L$ consistent with zero (i.e. no correlation). The lack of observed correlation between $Q_{\rm jet}$ and $L_\nu$, after accounting for the distance dependence, is due to the fact that the distance normalised range of both jet power and luminosity is very small relative to the intrinsic scatter and measurement uncertainty (see Figure \ref{fig:hotspots_data_plots}), leaving very little constraint on any intrinsic relationship after the effect of distance has been removed. The lack of observed correlation does not indicate the absence of an intrinsic relation, it merely indicates the shortcomings of this particular sample for the purpose of calibrating the scaling relation. 

\subsection{Minimum energy/spectral age jet power measurement}  \label{sec:daly_sample}

In a similar manner to the X-ray cavity systems, the time averaged jet power of an FRII radio galaxy may be estimated as $Q_{\rm jet} = 4pV/\tau$ \citep{leahy91}, where $p$ is the lobe internal pressure, $V$ is the lobe volume, and $\tau$ is the age of the source. Assuming cylindrical symmetry, and assuming that the lobe cross-section remains constant with time, \citet{wan00} estimate the rate of change of the volume as $V/\tau = \pi a_L^2 v_L$ where $a_L$ is the lobe cross-sectional radius, and $v_L$ is the rate of change of lobe length. \citet{wan00} then re-write the jet power as $Q_{\rm jet} = 4 \pi a_L^2 v_L p$. The lobe expansion rate $v_L$ is estimated from spectral ageing \citep{alexander87} using multifrequency radio observations. The lobe pressure $p$ is estimated from the radio luminosity, assuming magnetic field strength that is one-quarter of the minimum energy magnetic field strength \citep{miley80}. We note that the jet power derived in this way is not sensitive to offsets from minimum energy magnetic field strength \citep{odea09}. We also note that the expansion rate $v_L$ is likely to be systematically in error: spectral ages systematically underestimate dynamical ages by a significant factor, and have an uncertain relationship to the true source age \citep[eg.][]{eilek97, blundell00, kaiser05, hardcastle13b}.

Using the method of \citet{wan00}, \citet{daly12} present a compilation of jet power measurements for a sample of 31 3CRR powerful FRII radio galaxies, and use the observed correlation to determine the $Q_{\rm jet} - L_\nu$ scaling relation for powerful FRIIs. We note that this is the same method used by \citet{rawlings91} to estimate jet power, although in that case a different normalisation constant is assumed. 

We now wish to determine the distance dependence of the jet power measurements obtained in this way. We note that the lobe pressure is related to the minimum energy magnetic field strength as $p \propto B_{\rm me}^2$, and therefore
\begin{eqnarray}
Q \propto B_{\rm me}^2 a^2_L v_L.
\end{eqnarray}
Within this sample, the expansion velocity $v_L$ shows at best a modest positive correlation with redshift \citep[][Fig. 25]{odea09}, and covers a very narrow range, with a standard deviation that is only $60\%$ of the median value. Therefore, the main driving factors in the distance dependence are the minimum energy magnetic field strength and cross-section area of the lobes. As such, the distance dependence is the same as the hotspot method of GS13:
\begin{equation}
Q \propto D_L^{2 - \frac{2}{3 + \alpha}} \sim D_L^{1.5}
\end{equation}
Figure \ref{fig:daly_data_plots} shows the strong distance dependence of jet power in the sample \citep[see also][Figures 30 and 31]{odea09}. Partial correlation analysis on the sample of \citet{daly12} indicates that the observed correlation between $Q_{\rm jet}$ and $L_\nu$ is indeed dominated by the mutual distance dependence (Table 1), and this is confirmed by the results of our multivariable linear regression (Table 2). Following \citet{daly12}, we exclude Cygnus A from our analysis due to it being an extreme outlier relative to the linear model discussed here. 

The lack of observed correlation between $Q_{\rm jet}$ and $L_\nu$ in the sample of \citet{daly12}, is due to the fact that after statistically controlling for the distance dependence, the distance normalised range of both jet power and luminosity is very small relative to the intrinsic scatter and measurement uncertainty (see Figure \ref{fig:daly_data_plots}). After removing the effect of the mutual distance dependence, there is very little dynamic range in which to constrain the intrinsic relationship. As was the case with GS13, this is not an indication that no relation exists, it is merely an indication of the shortcomings of the sample for the purpose of calibrating the scaling relation.

\vspace{0.5cm}

\subsection{Emission line luminosity as a proxy for AGN jet power} \label{sec:emission_line_proxies}

In the preceding sections, we have shown that previously published calibrations of the $Q_{\rm jet} - L_\nu$ scaling relation for FRII radio galaxies are not reliable. In this section we consider the use of AGN emission line luminosities as a proxy for jet power as a means to calibrate the scaling relation for FRII radio galaxies. In particular, we highlight several issues that preclude an accurate calibration of the $Q_{\rm jet} - L_\nu$ scaling relation with this method.

\citet{willott99} determined an empirical relation between jet power and radio luminosity in FRII radio galaxies and quasars by treating the OII emission line luminosity as a proxy for jet mechanical power. \citet{willott99} based their analysis on two flux limited samples with significantly different flux limits (3CRR and 7C), and in doing so, were able to disentangle the effect of distance and intrinsic correlation. They found that the regression slope is $\beta_L = 0.79 \pm 0.04$, which agreed with their model prediction.

From a theoretical point of view, there are several caveats to bear in mind when treating emission line luminosity as a proxy for jet power. The emission line luminosity can only give an accurate calibration of the scaling relation if: (1) the emission line luminosity is linearly proportional to the photoionising luminosity of the accretion disk; (2) the photoionising luminosity from the disk is linearly proportional to accretion rate, and (3) the jet power is linearly proportional to the accretion power. It is not clear whether any of these conditions will be satisfied in reality, and the combination of all three is unlikely, as we now discuss.

Regarding the first condition, \citet{tadhunter98} performed photoionisation modelling to investigate the expected behaviour of various emission line fluxes with increasing photoionising luminosity. They find a relatively weak dependence of $L_{OII}$ on accretion power, and that the relationship differs significantly for OII and OIII emission lines. Furthermore, \citet{tadhunter98} find that the characteristics of the emission line clouds are not constant with radio power and/or redshift, indicating that any relationship between emission line luminosity and accretion power will be non-linear in general. 

The empirical relation between emission line and radio luminosity itself appears to be highly uncertain. \citet{willott99} find $L_{OII} \propto L_{151}^{0.79 \pm 0.04}$, while \citet{hardcastle09} find $L_{OII} \propto L_{178}^{1.02 \pm 0.2}$, and \citet{fernandes11} find $L_{OII} \propto L_{178}^{0.52 \pm 0.1}$.  We note that the difference between these results may be, at least partly, due to different regression methods used by each of the groups.

Further empirical uncertainty is observed when considering different proxies for the jet power.  \citet{hardcastle09} performed a comprehensive correlation analysis of the relationship between extended radio emission and accretion related AGN emission for a large sample of radio galaxies selected from the 3CRR catalogue, the majority of which are classed as FRII. Using this sample, they show that after accounting for the effect of distance, the total radio luminosity at 178 MHz is reasonably well correlated with several indicators of AGN accretion power such as X-ray, infrared and OIII narrow line luminosity. They do find, however, that the these correlations are non-linear, and the regression slopes are not consistent between the different proxies, with $0.7 < \beta_L < 1.4$ depending on which proxy is used. They argue that the best indicator of accretion power is the absorbed X-ray continuum luminosity, for which the regression slope is found to be $\beta_L = 0.72^{+0.12}_{-0.38}$. This result is formally consistent with the regression slope we find for FRI cavity systems. However, using the OIII line luminosity or infra-red luminosity, the regression slope is found to be $\beta_L \approx 1.4$. \citet{hardcastle09} argue that the OII line luminosity is not a good proxy for accretion related luminosity, since the correlation between OII line luminosity and the other indicators of accretion power are not significant after the common correlation with distance is accounted for.

In summary, we are unable to do determine an accurate calibration of the $Q_{\rm jet} - L_\nu$ scaling relation in FRII radio galaxies by using emission line luminosities as a proxy for jet power. We note, however, that the relation between accretion related AGN emission and radio luminosity cover a wide range of regression slopes, with $\beta_L \sim 0.5 - 1.4$. Bearing in mind the caveats described above, this may suggest a steeper relation between jet power and radio luminosity for FRIIs than we find for FRI X-ray cavity systems. As discussed in Section \ref{sec:model_predictions}, such a difference in the regression slope for FRI and FRII systems is expected, due to the difference in dynamics of the radio lobes.

\section{Comparison with model predictions}  \label{sec:model_predictions}

The relationship between jet power and radio luminosity depends largely on the source dynamics. \citet{willott99} developed a model-dependent predictor of jet power for FRII radio galaxies based on the self-similar model of FRII radio galaxy evolution \citep{kaiser97}. The self-similar model is applicable to powerful FRII radio galaxies in which the over-pressured lobes drive a strong bow-shock into the interstellar or intergalactic medium. In contrast, for X-ray cavity systems it is typically assumed that the expansion rate is subsonic, and is dictated by the buoyant velocity, which is not dependent on the jet power \citep{birzan04, birzan08, cavagnolo10, osullivan11,  mcnamara12}. The different dynamics of FRI and FRII radio galaxies implies a different scaling relation between jet power and radio luminosity, as we now show.

We begin with the following statement, which is the basis of X-ray cavity jet power measurements, and is equivalent to equation 4 of \citet{willott99}:
\begin{equation}
Q = \frac{H}{t} = \frac{\zeta p V}{t} 
\end{equation}
where H is the cavity enthalpy which accounts for the pV work done in expanding the cavity as well as the internal energy of the system. The pre-factor $\zeta$ depends on the equation of state of the plasma within the cavity, as well as the expansion history of the bubble \citep[eg.][]{mcnamara12}. It is often assumed that $\zeta = 4$, as appropriate for mature cavity systems in which the radio lobes are dominated by relativistic plasma \citep[][]{mcnamara12}, but $\zeta$ could be significantly higher in some cases if expansion is supersonic \citep[eg.][]{worrall12}. 

The pressure within the cavity is the sum of contributions from thermal particles $p_{\rm therm}$, relativistic particles $p_{\rm e}$, and magnetic field $p_B$: $p = p_{\rm therm} + p_{\rm e} + p_B$. We assume that the magnetic field is isotropically distributed (``tangled") on all scales, and therefore can be treated as a pressure term with effective pressure $p_B = (1/3)u_B$ \citep{leahy91}. In that case, the minimum pressure corresponds to minimum energy \citep{leahy91}, and we can write \citep{worrall06}
\begin{equation}
p_{\rm min} = \frac{1}{3}\frac{(\alpha+3)}{(\alpha+1)} \frac{B^2_{\rm mp}}{2 \mu_0}
\end{equation}
where $\alpha$ is the spectral index (defined such that $S_\nu \propto \nu^{-\alpha}$) and $B_{\rm mp}$ is the minimum pressure magnetic field strength given by 
\begin{equation}
B_{\rm mp} = \left[  \frac{(\alpha + 1)}{2} 2 \mu_0 C_1 \frac{(1+k_p)}{V} \frac{\gamma_{\rm min}^{1-2\alpha} \left(1 - \left( \frac{\gamma_{\rm max}}{\gamma_{\rm min}} \right)^{1-2\alpha} \right)} {(2\alpha - 1) } L_\nu \nu^\alpha \right]^{\frac{1}{\alpha + 3}}.
\end{equation}
Here $k_p$ is the ratio of pressure in thermal and relativistic particles, as opposed to $k$ -- the ratio of energy density in thermal and relativistic particles which is appropriate for the calculating the minimum energy magnetic field. $C_1$ is a function of $\alpha$, and involves several physical constants \citep{worrall06}. 

\newpage

Let us define $f = p/p_{\rm min}$, then ignoring physical constants and terms involving only the spectral index $\alpha$, we can write
\begin{eqnarray}
Q &=& \frac{\zeta f p_{\rm min} V}{t} \\
&\propto& \frac{\zeta f V^{1- \frac{2}{3+\alpha}} L_\nu^{\frac{2}{3+\alpha}} \gamma_{\rm min}^{\frac{2(1-2\alpha)}{3+\alpha}} (1+k_p)^{\frac{2}{(3+\alpha)}} }{t}  \label{eqn:Q_I}
\end{eqnarray}
Now let us parameterise the radio lobe dynamics as follows
\begin{eqnarray}
V &=& V_0 ~t^{n_t} ~Q^{n_Q} \label{eqn:n_V} \\
\Rightarrow t &\propto& V^{\frac{1}{n_t}} Q^{-\frac{n_Q}{n_t}}  \label{eqn:t}
\end{eqnarray}
Combining equations \ref{eqn:Q_I} and \ref{eqn:t} we can write
\begin{eqnarray}
Q &\propto& \left( \zeta f \right)^{\beta_{\zeta f}} V^{\beta_V} L_\nu^{\beta_{\rm L}} \gamma_{\rm min}^{\beta_{\gamma_{\rm min}}} (1+k_p)^{\beta_{k_p}}  \label{eqn:Q_II}
\end{eqnarray}
where, for $n_Q \neq n_t$:
\begin{eqnarray} 
\beta_{\zeta f} &=& \frac{1}{(1-\frac{n_Q}{n_t})} \label{eqn:n_params_1} \\
\beta_{V} &=& \frac{1 - \frac{2}{(3+\alpha)} - \frac{1}{n_t} }{(1-\frac{n_Q}{n_t})}  \\
\beta_{\rm L} &=& \frac{2}{(3 + \alpha)(1-\frac{n_Q}{n_t})} \label{eqn:beta_L}\\
\beta_{\gamma_{\rm min}} &=& \frac{2(1-2\alpha)}{(3 + \alpha)(1-\frac{n_Q}{n_t})}  \\
\beta_{k_p} &=& \frac{2}{(3 + \alpha)(1-\frac{n_Q}{n_t})} \label{eqn:n_params_2}
\end{eqnarray}
In the special case $n_Q = n_t$ the jet power is independent of luminosity (at a fixed volume), and in that case
\begin{eqnarray}
L_\nu \propto V^{  \frac{(3+\alpha)(1 - n_t)}{2 n_t} + 1}
\end{eqnarray}
This case may be of particular relevance to FRI radio galaxies, and is discussed further below.

\subsection{FRII lobe dynamics}

Models of FRII radio lobe dynamics have typically assumed that the lobe internal pressure is much greater than the external pressure of the ambient medium into which the lobes expand, resulting in supersonic, self-similar expansion \citep{begelman89, falle91, kaiser97}. If this is the case, the lobes evolve according to \citep{kaiser97, willott99}:
\begin{eqnarray}
V &\propto& t^{\frac{9}{5 - b}} Q^{\frac{3}{5 - b}}
\end{eqnarray}
so that $n_t = 9/(5-b)$ and $n_Q = 3/(5-b)$, where \textit{b} is the exponent of the power-law density profile $\rho \propto r^{-b}$. When substituted into Equations \ref{eqn:n_params_1} - \ref{eqn:n_params_2}, we obtain
\begin{eqnarray}  \label{eqn:FRII_scaling_relation}
Q \propto L_\nu^{\frac{3}{3+\alpha}} V^{ \left( \frac{4 + b}{6} - \frac{3}{3 + \alpha} \right) } (\zeta f)^{3/2}  (1+k_p)^{\frac{3}{3+\alpha}}  \gamma_{\rm min}^{\frac{3(1-2\alpha)}{3+\alpha}}
\end{eqnarray}
which is equivalent to the Willott relation for FRII radio galaxies \citep[see][]{osullivan11, shabala13}. For typical values of $\alpha \approx 0.8$, the exponent relating $Q_{\rm jet}$ and $L_\nu$ is $\beta_L({\rm FRII}) = 3/(3+\alpha) \approx 0.8$. 

We note, however, that the dynamics of FRII radio galaxy lobes remains a debated topic, and applicability of the self-similar dynamical model for FRII radio galaxies has been questioned, particularly for older sources. Estimates of the internal pressure of FRII radio galaxy lobes suggest that they may be close to pressure equilibrium with the external medium, rather than significantly over pressured as is required for supersonic, self-similar evolution \citep{hardcastle00, hardcastle02, croston04}. Furthermore, the distribution of axial ratios of FRII radio galaxies is dependent on linear size, with larger sources tending to have larger axial ratios \citep{mullin08}. This is at odds with self-similar models, in which the lobes remain over-pressured with respect to the external medium, and the axial ratio remains constant throughout the life of a source \citep{kaiser97}. We note that new dynamical models which include steepening of the gas density profiles with radius and thermal pressure of the ambient medium can explain the observed increase in the axial ratio of FRII radio galaxies \citep{turner15}. If the lobes of FRII radio galaxies do not evolve according to the self-similar models, the $Q_{\rm jet} - L_\nu$ scaling relation is expected to differ from that described by \citet{willott99}, and will depend on the values of $n_Q$ and $n_t$ as described by equations  \ref{eqn:Q_II} and \ref{eqn:beta_L}.

\subsection{FRI lobe dynamics}

The dynamics of FRI radio galaxy lobes may differ significantly for high and low power objects, and are likely to differ from that of powerful FRII radio galaxies. In this section, we discuss the variety of predictions for FRI lobe dynamics, and highlight the effect of the differing dynamics on the predicted $Q_{\rm jet} - L_\nu$ scaling relation.

Based on the torus-like appearance of the radio lobes of M87, \citet{churazov01} suggested that the lobe dynamics are dictated by buoyancy, in which case, the lobe expansion is independent of jet power. The assumption of buoyant bubble-like dynamics is often applied to X-ray cavity systems. While this is not true very early in the evolution of a radio source, it may be a good approximation for mature systems, like those that are preferentially detected in the X-ray cavities sample \citep{mcnamara12}, and therefore may provide a reasonable model for the evolution of the systems in question. However, see \citet{omma04} for an alternative view. Indeed, for FRI X-ray cavity systems, the source ages are often derived based on buoyant velocity estimates \citep[eg.][]{birzan04, cavagnolo10, osullivan11}. If the dynamics are dictated by buoyancy, the source evolution is independent of jet power, and $n_Q = 0$, in which case 
\begin{equation}
Q \propto L_\nu^{\frac{2}{3+\alpha}} V^{ \left(1 - \frac{2}{3 + \alpha}  - \frac{1}{n_t} \right) } \zeta f  (1+k_p)^\frac{2}{\left( 3 + \alpha \right)}  \gamma_{\rm min}^{\frac{2(1-2\alpha)}{\left( 3+\alpha \right) }}.
\end{equation}
For typical values of $\alpha \approx 0.8$, the exponent relating $Q_{\rm jet}$ and $L_\nu$ is $\beta_L({\rm FRI}) = 2/(3+\alpha) \approx 0.5$. 

Such a model may be valid for low power FRI radio galaxies. However, the evolution of more powerful FRI radio galaxies is unlikely to be well described by the buoyant bubble models. Powerful FRI X-ray cavity systems such as MS0735+7421 \citep{mcnamara05} and Hydra A \citep{nulsen05} show evidence for weak shocks surrounding the radio lobes, indicating the evolution of these systems is jet driven. In such systems, whilst the evolution will have some dependence on jet power, the dependence may be weaker than the case of highly supersonic, self-similar expansion, and as such, the implied $\beta_L({\rm FRI})$ may lie somewhere between that corresponding to a buoyantly rising bubble ($\beta_L \approx 0.5$) and that corresponding to supersonic, self-similar expansion ($\beta_L \approx 0.8$). 

Alternative models for FRI lobe dynamics have been proposed. For example \citet{luo10} present a so-called ``pressure-limiting" expansion model. In this model, the lobes evolve according to 
\begin{eqnarray}
V \propto t^{\frac{3}{3-b}} Q_{\rm jet}^{\frac{3}{3-b}}
\end{eqnarray}
where, again, $b$ is the exponent of the power-law density profile of the ambient medium. Therefore, in this model $n_t = n_Q$ and the jet power is independent of luminosity (for a given source size).

\subsection{Comparison of Predicted and Observed Scaling Relations in FRI X-ray cavity systems}  \label{sec:discussion_on_FRI_slope}

As shown in the preceding analysis, assuming $\alpha \lesssim 0.8$, the predicted regression slope is $| \beta_L | \gtrsim 0.5$ for all $n_Q \neq n_t$, and $\beta_L = 0$ for $n_Q = n_t$. Our empirically derived regression slope for the X-ray cavity sample ($\beta_L = 0.33 \pm 0.09$) is inconsistent with any model predictions: no dynamical model can produce such a flat regression slope. This implies that one or more of the additional model parameters ($V, \zeta, f, k_p, \gamma_{\rm min}$) are correlated with jet power in this sample in such a way that acts to flatten the observed regression slope, or the X-ray cavity jet powers contain a systematic bias, such that the jet power is underestimated in high power objects and/or jet power is overestimated in low power objects. We discuss each of these possibilities in more detail below.

We might expect that the value of $k_p$ (the ratio of pressure in non-radiating particles to the pressure in radiating particles) is anti-correlated with jet power, since weaker jets are likely to suffer more significant effects of entrainment \citep{bicknell97}. Indeed, within the FRI radio galaxies population, lower radio luminosities are typically associated with naked jet sources whilst the high luminosity end is dominated by lobed FRI radio galaxies \citep{parma02}. Naked jet sources typically have higher $k_p$ values than lobed FRIs \citep{croston08}. 

Additionally, unlike in the case of the self-similar dynamical model for FRII radio galaxies, the dependence of jet power on source size is not negligible. Considering the case of a buoyant bubble model, if the bubble expansion is adiabatic, and the bubble rises at a constant velocity in a power-law pressure profile of the form $p \propto R^{-b}$ then in Equation \ref{eqn:n_V} the exponent becomes $n_V = b/\gamma_{\rm ext} \sim 1$ where typically $b \approx 1 - 2$, and $\gamma_{\rm ext} = 5/3$ is the adiabatic index of the external medium, so that $\beta_V \approx -2/(3+\alpha)$. Source volumes in the X-ray cavities sample are correlated with jet power, due to the common distance dependence, and so this effect will tend to flatten the $Q_{\rm jet} - L_{\nu}$ scaling relation for this sample.

Finally, the discrepancy may be due to a systematic bias in the X-ray cavity power measurements. \citet{osullivan11} outline several sources of potential bias and uncertainty that could affect the accuracy of X-ray cavity estimates of jet power. Importantly, X-ray cavity powers neglect the energy associated with shocks. Shocks are likely to be more important for higher power objects \citep{mcnamara05, nulsen05, gitti10}, and therefore neglecting the energy in shocks may introduce a systematic bias in the X-ray cavity powers, underestimating the jet power of higher power sources, which would result in a flatter regression slope relative to the predicted value.

\section{Conclusions} \label{sec:conclusions}

What is the average relation (if any) between jet power and radio luminosity in radio galaxies? Do high and low power radio galaxies follow the same relation? We have addressed these questions from both an empirical and theoretical point of view. Our results may be summarised as follows:

\begin{enumerate}
\item The three methods previously used to calibrate the $Q_{\rm jet} - L_\nu$ scaling relation have a strong dependence on distance, each with $Q_{\rm jet} \sim D_L^{1.5}$. The mutual distance dependence induces a spurious relation between $Q_{\rm jet}$ and $L_\nu$. The similar distance dependence for each of the jet power measurement techniques accounts for the apparent similarity of previously reported scaling relations for FRI and FRII radio galaxies. 
\item For FRI X-ray cavity systems, after accounting for the mutual distance dependence, we find a very weak correlation between jet power and radio luminosity, with $\log(Q_{\rm jet}) \propto \beta_L \log(L_\nu)$ where $\beta_L = 0.33 \pm 0.09$ at 327~MHz and $\beta_L = 0.27 \pm 0.09$ at 1400~MHz. 
\item For powerful FRII radio galaxies, after accounting for the mutual distance dependence, we find no evidence for an intrinsic relationship between $Q_{\rm jet}$ and $L_\nu$. However, the lack of evidence of an intrinsic correlation does not necessarily imply that there is no intrinsic correlation, simply that the samples do not span enough range in distance-normalised parameter space to accurately constrain the intrinsic relation. 
\item Proxies for jet power such as the X-ray, infra-red or narrow emission line luminosities of the AGN, indicate that the scaling relation for FRII radio galaxies may be significantly steeper than that obtained for FRIs. However, the uncertain non-linear relationships between accretion related emission and jet power means that an accurate empirical calibration of the scaling relation for FRIIs is not possible with this approach. The broad range in regression slopes obtained when using different proxies for jet power demonstrates the difficulties faced in using AGN related emission to calibrate the scaling relation. 
\item From a theoretical point of view, we have shown that the different dynamics of FRI and FRII radio galaxy lobes implies a difference in the expected $Q_{\rm jet} - L_\nu$ scaling relation. Taking the common assumption that the dynamics of X-ray cavity systems are similar to a buoyantly rising bubble, such that the evolution is dictated by the external environment, we predict for these systems $\log(Q_{\rm jet}) \sim 0.5 \log(L_\nu)$. In contrast, FRII systems evolve on a jet driven timescale, which results in a prediction of $\log(Q_{\rm jet}) \sim 0.8 \log(L_\nu)$, as first described by \citet{willott99}. The flatter slope for FRI radio galaxies relative to FRII radio galaxies is consistent with our conclusion in point (iv). We note however that to achieve a flat regression slope $\beta_L < 0.5$, additional model parameters must be correlated with jet power in such a way as to flatten the observed scaling relation, or there must be a systematic bias in X-ray cavity jet powers such that the jet power is underestimated in high luminosity objects and/or jet power is overestimated in low luminosity objects. 
\item Finally, we repeat the arguments of GS13 regarding the relative scaling relations of FRI and FRII radio galaxies. Results from the analysis of radio galaxies and their hot X-ray emitting atmospheres suggest that non-radiating particles dominate the energy budget in the lobes of FR I radio galaxies, in some cases by a factor of more than 1000 \citep{croston03, croston08, birzan08}, while radiating particles dominate the energy budget in FR II radio galaxy lobes \citep{croston04, croston05, belsole07}. This implies a significant difference in the radiative efficiency of the two morphological classes, which would manifest as a large difference in the normalisation of the $Q_{\rm jet} - L_\nu$ scaling relations. 
\end{enumerate}

We conclude that the $Q_{\rm jet} - L_\nu$ scaling relations remain poorly constrained through observations. Furthermore, the uncertainty regarding radio lobe dynamics provides some uncertainty in the predicted scaling relations. However, our analysis indicates that FRI and FRII radio galaxies do not follow the same scaling relation between jet power and radio luminosity: the regression slope for FRI X-ray cavity systems is significantly flatter than previously reported, with $\log(Q_{\rm jet}) \propto (0.33 \pm 0.09) \log(L_\nu)$ at 327 MHz. This revision in the scaling relation gives a greater energetic importance to low luminosity radio galaxies, which has strong implications for studies of radio mode feedback. Low luminosity radio galaxies typically deposit energy at smaller radii than more powerful systems, because they do not expand to 100 kpc sizes. As such, low power radio galaxies may be more effective at offsetting cooling in hot atmospheres of massive galaxies, groups and clusters by depositing more energy in the regions where it is most needed to offset cooling. \\

Dynamical models \citep[eg.][]{turner15} and simulations \citep[eg.][]{hardcastle13, hardcastle14} will help to predict more accurately the $Q_{\rm jet} - L_\nu$ relation as a function of the radio source morphology, environment, cosmic epoch, and host galaxy history. This will provide a framework for interpreting data for next-generation continuum surveys from LOFAR, ASKAP, MWA, MeerKAT and the Square Kilometre Array.

\section*{Acknowledgements}

The authors thank Dave Jauncey for insightful conversations at the outset of this work, and the anonymous referee for constructive feedback. 

The research leading to these results has received funding from the European Research Council under the European Union's Seventh Framework Programme (FP/2007-2013) / ERC Advanced Grant RADIOLIFE-320745. 

SSS thanks the Australian Research Council for an Early Career Fellowship (DE130101399).

{}


\end{document}